\documentclass[fleqn]{article}
\usepackage{amssymb,latexsym}
\usepackage{amsmath, amsthm}
\usepackage{color}
\usepackage[dvips]{graphicx}
\newtheorem{Theorem}{Theorem}
\newtheorem{theo}{Proposition}
\newtheorem{lemma}{Lemma}

{\catcode `\@=11 \global\let\AddToReset=\@addtoreset}
\AddToReset{lemma}{section}

\newcommand{\sgn}{\mathop {\rm sgn}\nolimits}
\newcommand{\abar}{\chi}
 \newcommand{\tQ}{{\widetilde Q}}

\newcommand{\be}{\begin{equation}}
\newcommand{\ee}{\end{equation}}
\newcommand{\SC}{S_{\times}}
\newcommand{\rd}{{\rm d}} 
\newcommand{\iv}{\raisebox{-2.15pt}{$\lrcorner$}\hspace*{-3.3pt}\rfloor}
\begin{document}

\title{\bf Geometry of crossing null shells}

\author{Jacek Jezierski
\\ Katedra Metod Matematycznych Fizyki, \\ ul. Ho\.{z}a 74,
00-682 Warszawa, Poland}

\date{\small 04.20.Fy, 04.40.-b, 04.60.Ds}
\maketitle

\begin{abstract}
New geometric objects on null thin layers are introduced and their
importance for crossing null-like shells are discussed. The
Barrab\`es--Israel equations are represented in a new geometric
form and they split into decoupled system of equations for two
different geometric objects: tensor density ${\bf G}^a{_b}$ and
vector field $I$. Continuity properties of these objects through a
crossing sphere are proved. In the case of spherical symmetry
Dray--t'Hooft--Redmount formula results from continuity property
of the corresponding object.
\end{abstract}



\section{Introduction}
 Self gravitating matter shell (see \cite{shell-massive,shell1a})
 became an important laboratory for testing global
properties of gravitational field interacting with matter. Models
of a thin matter layer allow us to construct useful
mini-superspace examples. Toy models of quantum gravity, started
by Dirac {\cite{Dirac}, may give us a deeper insight into a possible future
shape of the quantum theory of gravity (see \cite{shell-quantum,Honnef}).
Especially interesting are null-like shells, carrying a
self-gravitating light-like matter
(see \cite{hajicek-null,H-Kou1,H-Kou2,H-Kou3}).
Classical equations of motion of such a shell have been derived by
Barrab\`es and Israel in their seminal paper \cite{IB}.
Junction conditions for general hypersurfaces in spacetime
are also given in \cite{MarSen}.

A complete Lagrangian and Hamiltonian description of the theory of
self-gra\-vi\-ta\-ting light-like matter shell,
which is no longer spherically symmetric,
 was given (in terms
of gauge-independent geometric quantities) in \cite{jjPRD02}.
For this purpose the
notion of an extrinsic curvature for a null-like hypersurface was
discussed and the corresponding Gauss--Codazzi equations were
proved. These equations imply Bianchi identities for spacetimes
with null-like, singular curvature. Energy-momentum tensor density
of a light-like matter shell is unambiguously defined in terms of
an invariant matter Lagrangian density. Noether identity and
Belinfante--Rosenfeld theorem for such a tensor density was also proved.
Finally, the Hamiltonian dynamics of the interacting system:
``gravity + matter'' was derived from the total Lagrangian, the
latter being an invariant scalar density.

Starting from the action functional for a single spherical shell due to
Louko, Whiting and Friedman \cite{LWF}, H\'{a}j\'{\i}\v{c}ek and Kouletsis
generalized it for any number of spherically symmetric null shells,
including the cases, when the shells intersect \cite{H-Kou2}.

In this paper we consider a general non-symmetric
case of two crossing null shells. It occurs that the geometric
objects on the null shells are continuous through an intersecting
sphere due to the observation that ``jump of the jump'' vanishes
(see Lemma \ref{jjf}). This implies that the dynamics of the crossing
shells is described by the equations for a single shell plus
continuity property across intersecting sphere.

 We also discuss
a special case of spherical symmetry. In particular,
we give a simple argument (in the case of spherical symmetry)
 for triviality of
 the whole ``ADM-momentum'' tensor density ${\bf G}^a{_b}$
  which implies that
 the corresponding energy-momentum tensor density $\tau^a{_b}$
of a light-like matter shell is vanishing.

Geometry of a single shell introduced in \cite{jjPRD02}
is completed by an extra object --- a null vector field $I$,
which is always well defined on a null shell
and does not vanish in the case of spherical symmetry.
Roughly speaking, in the case of a null shell
 the ``ADM-momentum'' tensor density ${\bf G}^{ab}$
(which is well defined for any non-degenerate surface $S$)
 splits into two geometric
objects: a tensor density ${\bf G}^{a}{_b}$ and a null vector
$I^a$. They contain a similar information as the jump of a
``transverse'' extrinsic curvature ${\cal K}_{ab}$ in
Barrab\`es--Israel approach.

The dynamical system constituted of two spherically symmetric
 null shells has been studied in \cite{H-Kou1}. The shells at
 intersection sphere $\SC$
 exchange energy according to the
 Dray--t'Hooft--Redmount formula \cite{DH,Red}.
We show that the continuity of the metric (around intersection
sphere) implies the continuity of the vector field $I$ through
$\SC$ on both shells. Moreover,
in the case of spherical symmetry we show that the continuity
of $I$ gives the Dray--t'Hooft--Redmount formula.
This means that our new object should be useful in
 generalizations of the Dray--t'Hooft--Redmount formula
 for the case of crossing two null shells without any symmetry.

\section{Geometry of a single null shell}
\subsection{Geometry of a null hypersurface and Gauss--Codazzi
constraints} \label{geometry1}

A null hypersurface in a Lorentzian spacetime $M$ is a
three-dimensional submanifold $S \subset M$ such that the
restriction $g_{ab}$ of the spacetime metric $g_{\mu\nu}$ to $S$
is degenerate.

We shall often use adapted coordinates, where coordinate $x^3$ is
constant on $S$. Space coordinates will be labeled by $k,l =
1,2,3$; coordinates on $S$ will be labeled by $a,b=0,1,2$;
finally, coordinates on $S_{t} := V_{t} \cap S$ (where $V_{t}$ is
a Cauchy surface corresponding to constant value of the
``time-like'' coordinate $x^0=t$) will be labeled by $A,B=1,2$.
Spacetime coordinates will be labeled by Greek characters
$\alpha, \beta, \mu, \nu$.

The non-degeneracy of the spacetime metric implies that the
metric $g_{ab}$ induced on $S$ from the spacetime metric
$g_{\mu\nu}$ has signature $(0,+,+)$. This means that there is a
non-vanishing null-like vector field $X^a$ on $S$, such that its
four-dimensional embedding $X^\mu$ to $M$ (in adapted coordinates
$X^3=0$) is orthogonal to $S$. Hence, the covector $X_\nu = X^\mu
g_{\mu\nu} = X^a g_{a\nu}$ vanishes on vectors tangent to $S$ and,
therefore, the following identity holds:
\begin{equation}\label{degeneracy}
  X^a g_{ab} \equiv 0 \ .
\end{equation}
It is easy to prove (cf. \cite{JKC}) that integral curves of
$X^a$, after a suitable reparameterization, are geodesic curves of
the spacetime metric $g_{\mu\nu}$. Moreover, any null
hypersurface $S$ may always be embedded in a one-parameter
congruence of null hypersurfaces.

We assume that topologically we have $S = {\mathbb R}^1 \times
S^2$. Since our considerations are purely local, we fix the
orientation of the ${\mathbb R}^1$ component and assume that
null-like vectors $X$ describing degeneracy of the metric $g_{ab}$ of
$S$ will be always compatible with this orientation. Moreover, we
shall always use coordinates such that the coordinate $x^0$
increases in the direction of $X$, i.e.,~inequality $X(x^0) = X^0
> 0$ holds. In these coordinates degeneracy fields are of the form
$X = f(\partial_0-n^A\partial_A)$, where $f > 0$, $n_A = g_{0A}$
and we rise indices with the help of the two-dimensional matrix
${\tilde{\tilde g}}^{AB}$, inverse to $g_{AB}$.

If by $\lambda$ we denote the two-dimensional volume form on each
surface $x^0 = \mbox{\rm const.}$:
\begin{equation}\label{lambda}
  \lambda:=\sqrt{\det g_{AB}} \ ,
\end{equation}
then for any degeneracy field $X$ of $g_{ab}$ the following
object
\[
v_{X} := \frac {\lambda}{X(x^0)}
\]
is a well defined scalar density on $S$
according to \cite{jjPRD02}.  This means
that ${\bf v}_X := v_X dx^0 \wedge dx^1 \wedge dx^2$ is a
coordinate-independent differential three-form on $S$. However,
$v_X$ depends upon the choice of the field  $X$.

It follows immediately from the above definition that the
following object:
\[
\Lambda = v_X \ X
\]
is a well defined (i.e.,~coordinate-independent) vector density on
$S$. Obviously, it {\em does not depend} upon any choice of the
field $X$:
\begin{equation}\label{Lambda}
\Lambda =  \lambda (\partial_0-n^A\partial_A) \ .
\end{equation}
Hence, it is an intrinsic property of the internal geometry
$g_{ab}$ of $S$. The same is true for the divergence $\partial_a
\Lambda^a$, which is, therefore, an invariant, $X$-independent,
scalar density on $S$. Mathematically (in terms of differential
forms), the quantity $\Lambda$ represents the two-form:
\[
{\bf L} := \Lambda^a \left( \partial_a \; \iv  \;  dx^0 \wedge
dx^1 \wedge dx^2 \right) \ ,
\]
whereas the divergence represents its exterior derivative (a
three-from): d${\bf L} := \left( \partial_a \Lambda^a \right)dx^0
\wedge dx^1 \wedge dx^2$. In particular, a null surface with
vanishing d${\bf L}$ is called a {\em non-expanding horizon} (see
\cite{AL}).

Both objects ${\bf L}$ and ${\bf v}_X$ may be defined
geometrically, without any use of coordinates. For this purpose we
note that at each point $x \in S$, the tangent space $T_xS$ may be
quotiented with respect to the degeneracy subspace spanned by $X$.
The quotient space carries a non-degenerate Riemannian metric and,
therefore, is equipped with a volume form $\omega$ (its coordinate
expression would be: $\omega = \lambda \ dx^1 \wedge dx^2$). The
two-form ${\bf L}$ is equal to the pull-back of $\omega$ from the
quotient space to $T_xS$. The three-form ${\bf v}_X$ may be
defined as a product: ${\bf v}_X = \alpha \wedge {\bf L}$, where
$\alpha$ is {\em any} one-form on $S$, such that $<X,\alpha>
\equiv 1$.

The degenerate metric $g_{ab}$ on $S$ does not allow to define
{\em via} the compatibility condition $\nabla g = 0$, any natural
connection, which could be applied to generic tensor fields on $S$.
Nevertheless, there is one exception: it was shown in \cite{jjPRD02}
that the degenerate metric defines {\em uniquely} a certain covariant,
first order differential operator.
 The operator may be applied only to mixed
(contravariant-covariant) tensor density fields ${\bf H}^{a}{_b}$,
satisfying the following algebraic identities:
\begin{eqnarray}
{\bf H}^{a}{_b} X^b = 0 \ , \label{G-1} \\ {\bf H}_{ab} = {\bf
H}_{ba} \ , \label{G-2}
\end{eqnarray}
where ${\bf H}_{ab} := g_{ac} {\bf H}^{c}{_b}$. Its definition
cannot be extended to other tensorial fields on $S$. Fortunately,
 the extrinsic curvature of a null-like surface and
the energy-momentum tensor of a null-like shell are described by
tensor densities of this type.

The operator, which we denote by ${\overline{\nabla} }_a$,
is defined by means of the four-dimensional
metric connection in the ambient spacetime $M$ in the following
way: Given ${\bf H}^{a}_{\ b}$, take any its extension ${\bf
H}^{\mu\nu}$  to a four-dimensional, symmetric tensor density,
``orthogonal'' to $S$, i.e. satisfying ${\bf H}^{\perp\nu}=0$
(``$\perp$'' denotes the component transversal to $S$). Define
${\overline{\nabla} }_a {\bf H}^{a}_{\ b}$ as the restriction to
$S$ of the four-dimensional covariant divergence ${\nabla}_\mu
{\bf H}^{\mu}_{\ \nu}$.
It was shown in \cite{jjPRD02} that ambiguities
which arise when extending three-dimensional object ${\bf
H}^{a}_{\ b}$ living on $S$ to the four-dimensional one, cancel
finally and the result is unambiguously defined as a
covector density on $S$. It turns out, however, that this result
does not depend upon the spacetime geometry and may be defined
intrinsically on $S$ as follows:
\begin{eqnarray}
    \nonumber
    {\nabla}_a {\bf H}^{a}_{\ b} & = &
     \label{covariant-deg}
     \partial_a {\bf H}^{a}_{\ b} -
    \frac 12 {\bf H}^{ac} g_{ac , b} \ ,
\end{eqnarray}
where $g_{ac , b} := \partial_b g_{ac}$,
 a tensor density ${\bf
H}^{a}_{\ b}$ satisfies identities (\ref{G-1}) and (\ref{G-2}),
and moreover, ${\bf H}^{ac}$ is {\em any} symmetric tensor
density, which reproduces ${\bf H}^{a}_{\ b}$ when lowering an
index:
\begin{equation}\label{G-mixed}
  {\bf H}^{a}_{\ b}  = {\bf H}^{ac} g_{cb} \, .
\end{equation}
It is easily seen, that such a tensor density always exists due to
identities (\ref{G-1}) and (\ref{G-2}), but the reconstruction of
${\bf H}^{ac}$ from ${\bf H}^{a}_{\ b}$ is not unique, because
${\bf H}^{ac} + C X^a X^c$ also satisfies (\ref{G-mixed}) if ${\bf
H}^{ac}$ does. Conversely, two such symmetric tensors ${\bf
H}^{ac}$ satisfying (\ref{G-mixed}) may differ only by $C X^a
X^c$. Fortunately, this non-uniqueness does not influence the
value of (\ref{covariant-deg}). Hence, the following definition
makes sense:
\begin{equation}\label{div-final}
  {\overline{\nabla} }_a {\bf H}^{a}_{\ b}  :=
\partial_a {\bf H}^{a}_{\ b} - \frac 12 {\bf H}^{ac}
g_{ac , b} \ .
\end{equation}
The right-hand-side does not depend upon any choice of coordinates
(i.e.,~transforms like a genuine covector density under change of
coordinates).

To express directly the result in terms of the original tensor
density ${\bf H}^{a}_{\ b}$, we observe that it has five
independent components and may be uniquely reconstructed from
${\bf H}^{0}_{\ A}$ (2 independent components) and the symmetric
two-dimensional matrix ${\bf H}_{AB}$ (3 independent components).
Indeed, identities (\ref{G-1}) and (\ref{G-2}) may be rewritten as
follows:
\begin{align}
{\bf H}^{A}_{\ B} & =  {\tilde{\tilde g}}^{AC}{\bf H}_{CB} - n^A
{\bf H}^{0}_{\ B} \ , \label{AB}
\\ {\bf H}^{0}_{\ 0} & =  {\bf H}^{0}_{\ A} n^A \ ,  \label{00}
\\ {\bf H}^{B}_{\ 0} & =  \left( {\tilde{\tilde g}}^{BC}{\bf H}_{CA}
- n^B {\bf H}^{0}_{\ A} \right) n^A \label{B0} \ .
\end{align}
The correspondence between ${\bf H}^{a}{_b}$ and $({\bf
H}^{0}{_A},{\bf H}_{AB})$ is one-to-one.

To reconstruct ${\bf H}^{ab}$  from ${\bf H}^{a}{_b}$ up to an
arbitrary additive term $C X^a X^b$, take the following,
coordinate dependent, symmetric quantity:
\begin{align}
{\bf F}^{AB} & :=  {\tilde{\tilde g}}^{AC} {\bf H}_{CD }
{\tilde{\tilde g}}^{DB} - n^A {\bf H}^{0}_{\ C } {\tilde{\tilde
g}}^{CB} - n^B {\bf H}^{0}_{\ C } {\tilde{\tilde g}}^{CA} \ ,
\\ {\bf F}^{0A} & :=  {\bf H}^{0}_{\ C } {\tilde{\tilde g}}^{CA}
=: {\bf F}^{A0} \ ,
\\ {\bf F}^{00} & :=  0 \ .
\end{align}
It is easy to observe that any ${\bf H}^{ab}$ satisfying
(\ref{G-mixed}) must be of the form:
\begin{equation}\label{reconstr}
{\bf H}^{ab} = {\bf F}^{ab} + {\bf H}^{00} X^a X^b \ .
\end{equation}
The non-uniqueness in the reconstruction of ${\bf H}^{ab}$ is,
therefore, completely described by the arbitrariness in the choice
of the value of ${\bf H}^{00}$. Using these results, we finally
obtain:
\begin{eqnarray}
{\overline{\nabla} }_a {\bf H}^{a}_{\ b} & := &
\partial_a {\bf H}^{a}_{\ b} - \frac 12 {\bf H}^{ac}
g_{ac , b} =
\partial_a {\bf H}^{a}_{\ b} - \frac 12 {\bf F}^{ac}
g_{ac , b}\nonumber
\\  & = & \hspace*{-1ex} \partial_a {\bf H}^{a}_{\ b} - \frac 12 \left(
2 {\bf H}^{0}_{\ A} \ n^A_{ \ ,b} - {\bf H}_{AC}  {\tilde{\tilde
g}}^{AC}_{\ \ \ ,b} \right) \, . \label{Coda}
\end{eqnarray}
The operator on the right-hand-side of (\ref{Coda}) is
called the (three-dimen\-sio\-nal) covariant derivative of ${\bf
H}^{a}_{\ b}$ on $S$ with respect to its degenerate metric
$g_{ab}$. It was proved in \cite{jjPRD02} that it is well defined
(i.e.,~coordinate-independent) for a tensor density ${\bf
H}^{a}_{\ b}$ fulfilling conditions (\ref{G-1}) and (\ref{G-2}).
It was also shown
 that the above definition coincides with the one given in
terms of the four-dimensional metric connection and
due to (\ref{covariant-deg}), it equals:
\begin{equation}\label{int-ext}
  \nabla_\mu {\bf H}^{\mu}_{\ b} =
  \partial_\mu {\bf H}^{\mu}_{\ b} -
    \frac 12 {\bf H}^{\mu\lambda} g_{\mu\lambda , b} =
    \partial_a {\bf H}^{a}_{\ b} -
    \frac 12 {\bf H}^{ac} g_{ac , b} \ ,
\end{equation}
and, whence, coincides with ${\overline{\nabla} }_a {\bf H}^{a}_{\
b}$ defined intrinsically on $S$.


To describe exterior geometry of $S$ we begin with covariant
derivatives {\em along} $S$ of the ``orthogonal vector $X$''.
Consider the tensor $\nabla_a X^\mu$. Unlike in the non-degenerate
case, there is no unique ``normalization'' of $X$ and, therefore,
such an object does depend upon a choice of the field $X$. The
length of $X$  vanishes. Hence, the tensor is
again orthogonal to $S$, i.e.,~the components corresponding to
$\mu = 3$ vanish identically in adapted coordinates. This means
that $\nabla_a X^b$ is a purely three-dimensional tensor living on
$S$. For our purposes it is useful to use the ``ADM-momentum'' version
of this object, defined in the following way:
\begin{equation}\label{Q-fund}
{Q^a}_b (X) := -s \left\{ v_X \left( \nabla_b X^a - \delta_b^a
\nabla_c X^c \right) + \delta_b^a \partial_c \Lambda^c \right\}
\, ,
\end{equation}
where $s:=\sgn g^{03}=\pm 1$. Due to above convention, the
object ${Q^a}_b (X)$ feels only {\em external
orientation} of $S$ and does not feel any internal orientation of
the field $X$.

\underline{Remark:} If $S$ is a {\em non-expanding horizon}, the last
term in the above definition vanishes.

The last term in (\ref{Q-fund}) is $X$-independent. It has been
introduced in order to correct algebraic properties of the
quantity $ v_X \left( \nabla_b X^a - \delta_b^a \nabla_c X^c
\right)$: it was shown in \cite{jjPRD02} that
${Q^a}_b$ satisfies identities
(\ref{G-1})--(\ref{G-2}) and, therefore, its covariant divergence
with respect to the degenerate metric $g_{ab}$ on $S$ is uniquely
defined. This divergence enters into the Gauss--Codazzi equations,
which
relate the divergence of $Q$ with the transversal component ${\cal
G}^{\perp}_{\ b}$ of the Einstein tensor density ${\cal G}^\mu_{\
\nu} = \sqrt{|\det g |} \left( R^\mu_{\ \nu} - \delta^\mu_\nu
\frac 12 R \right)$. The transversal component of such a
tensor density is a well defined three-dimensional object living
on $S$. In coordinate system adapted to $S$, i.e.,~such that the
coordinate $x^3$ is constant on $S$, we have ${\cal G}^{\perp}_{\
b} = {\cal G}^{3}_{\ b}$. Due to the fact that ${\cal G}$ is a
tensor density, components ${\cal G}^{3}_{\ b}$ {\em do not
change} with changes of the coordinate $x^3$, provided it remains
constant on $S$. These components describe, therefore, an
intrinsic covector density living on $S$.
\begin{theo}
The following null-like-surface version of the Gauss--Codazzi
equation is true:
\begin{equation}\label{G-C}
    {\overline{\nabla} }_a {Q}^{a}_{\ b}(X) +s v_{X} \partial_b
    \left( \frac {\partial_c \Lambda^c}{v_{X}} \right) \equiv
    -{\cal G}^{\perp}_{\ b}  \ .
\end{equation}
 \end{theo}
We remind the reader that the ratio between two scalar densities:
$\partial_c \Lambda^c$ and $v_X$, is a scalar function. Its
gradient is a covector field. Finally, multiplied by the density
$v_X$, it produces an intrinsic covector density on $S$. This
proves that also the left-hand-side is a well defined geometric
object living on $S$. The equation (\ref{G-C}) is closely related
to Raychaudhuri \cite{Raych} equation for the congruence of
null geodesics generated by the vector field $X$.

\subsection{Bianchi identities for spacetimes with distribution
valued curvature} \label{Bianchi}

In this paper we consider a spacetime $M$ with distribution
valued curvature tensor in the sense of Taub \cite{Taub}. This
means that the metric tensor, although continuous, is not
necessarily $C^1$-smooth across $S$: we assume that the connection
coefficients $\Gamma^\lambda_{\mu\nu}$ may have only step
discontinuities (jumps) across $S$. Formally, we may calculate the
Riemann curvature tensor of such a spacetime, but derivatives of
these discontinuities with respect to the variable $x^3$ produce a
$\boldsymbol\delta$-like, singular part of $R$:
\begin{equation}\label{R-sing}
  \mbox{\rm sing}{(R)^\lambda}_{\mu\nu\kappa} = \left( \delta^3_\nu
[ \Gamma^\lambda_{\mu\kappa} ] - \delta^3_\kappa [
\Gamma^\lambda_{\mu\nu} ] \right) \boldsymbol\delta (x^3) \ ,
\end{equation}
where by $\boldsymbol\delta$ we denote the Dirac distribution (in
order to distinguish it from the Kronecker symbol $\delta$) and by
$[f]$ we denote the jump of a discontinuous quantity $f$ between
the two sides of $S$. The above formula is invariant under {\em
smooth} transformations of coordinates. There is, however, no
sense to impose such a smoothness across $S$. In fact, the
smoothness of spacetime is an independent condition on both sides
of $S$. The only reasonable assumption imposed on the
differentiable structure of $M$ is that the metric tensor --- which
is smooth separately on both sides of $S$ --- remains
continuous across $S$.
Admitting coordinate transformations preserving the above
condition, we loose a  part of information contained in quantity
(\ref{R-sing}), which becomes now coordinate-dependent. It turns
out, however, that another part, namely the Einstein
tensor density calculated from (\ref{R-sing}), preserves its
geometric, intrinsic (i.e.,~coordinate-independent) meaning. In
case of a non-degenerate geometry of $S$, the following formula
was used by many authors (see \cite{shell-massive, shell1a,
shell-quantum,shell1,shell1b}):
\begin{equation}\label{E-sing-non-deg}
\mbox{\rm sing}({\cal G})^{\mu\nu }  =   {\bf G}^{\mu\nu}
\boldsymbol\delta(x^3) \ ,
\end{equation}
where the ``transversal-to-$S$'' part of ${\bf G}^{\mu\nu}$
vanishes identically:
\begin{equation}\label{G^perp=0}
{\bf G}^{\perp \nu} \equiv 0 \ ,
\end{equation}
and the ``tangent-to-$S$'' part ${\bf G}^{ab}$ equals to the jump
of the ADM-momentum\footnote{$Q^{ab}=
\sqrt{|\det g_{cd}|}(g^{ab}\mathrm{tr}K-K^{ab})$,
where $g^{ab}$ is the inverse three-metric and
$K^{ab}$ is an extrinsic curvature.}  $Q^{ab}$ of $S$ between the two
sides of the surface:
\begin{equation}\label{G-Q}
  {\bf G}^{ab} = [Q^{ab}] \ .
\end{equation}
This quantity is a purely {\em three-dimensional}, symmetric
tensor density living on $S$. When multiplied by the {\em
one-dimensional} density $\boldsymbol\delta(x^3)$ in the
transversal direction, it produces the {\em four-dimensional}
tensor density ${\cal G}$ according to formula
(\ref{E-sing-non-deg}).

In the case of our degenerate surface $S$ it was shown in
\cite{jjPRD02} that formulae
(\ref{E-sing-non-deg}) and (\ref{G^perp=0}) remain valid also in
this case. In particular, the latter formula means that the
four-dimensional quantity ${\cal G}^{\mu\nu }$ reduces in fact to
an intrinsic, three-dimensional quantity living on $S$. However,
formula (\ref{G-Q}) cannot be true, because --- as we have
seen --- there is no way to define uniquely the object $Q^{ab}$ for
the degenerate metric on $S$. Instead, we are able to prove the
following formula:
\begin{equation}\label{Einst-Q}
  {{\bf G}^a}_b = [ {Q}^{a}_{\ b}(X) ] \ ,
\end{equation}
where the bracket denotes the jump of ${Q}^{a}_{\ b}(X)$ between
the two sides of the singular surface. This quantity
{\em does not depend} upon any choice of $X$ and
 the singular part $\mbox{\rm sing}{({\cal G})^a}_b$ of
the Einstein tensor is well defined. We will show in the sequel
that the missing component ${\bf G}^{00}$ can be recovered in another
geometric object, which is presented in the next Section.

\underline{Remark:} Otherwise as in the non-degenerate case, the
contravariant components ${\bf G}^{ab}$ in formula
(\ref{E-sing-non-deg}) do not transform as a tensor density on
$S$. Hence, the quantity defined by these components would be
coordinate-dependent. According to (\ref{Einst-Q}), ${\bf G}$
becomes an intrinsic three-dimensional tensor density on $S$ only
after lowering an index, i.e.,~in the version of ${{\bf G}^a}_b$.
This proves that ${\bf G}^{\mu\nu}$ may be reconstructed from
${{\bf G}^a}_b$ up to an additive term $C X^\mu X^\nu$ only. We
stress that the dynamics of the shell
 is unambiguously expressed in terms of the
gauge-invariant, intrinsic quantity ${{\bf G}^a}_b$.


We conclude that the total Einstein tensor of our spacetime is a
sum of the regular part\footnote{The regular part is a smooth
tensor density on both sides of the surface $S$ (calculated for
the metric $g$ separately) with possible step discontinuity across
$S$.} $\mbox{\rm reg}({\cal G})$ and the above singular part
$\mbox{\rm sing}({\cal G})$ living on the singularity surface $S$.
Thus
\begin{equation}\label{Bianchi-1}
  {{\cal G}^\mu}_\nu = \mbox{\rm reg}{({\cal G})^\mu}_\nu +
  \mbox{\rm sing}{({\cal G})^\mu}_\nu \ ,
\end{equation}
and the singular part is given {\em up to an additive term} $C
X^\mu X_\nu \boldsymbol\delta(x^3)$. The
following {\em four-dimensional} covariant divergence is
unambiguously defined:
\begin{equation}
0 = \nabla_\mu {\cal G}^{\mu}_{\ c} =
\partial_\mu {\cal G}^{\mu}_{\ c}
- {\cal G}^{\mu}_{\ \alpha} \Gamma^\alpha_{\mu c} = \partial_\mu
{\cal G}^{\mu}_{\ c} - \frac 12 {\cal G}^{\mu\lambda}
g_{\mu\lambda , c} \ .\label{bian-ident}
\end{equation}
It is proved in \cite{jjPRD02} that this quantity vanishes identically
and the total singular part of the Bianchi identities
reads:
\begin{equation}\label{Bianchi-fund}
    \mbox{\rm sing}\left(  \nabla_\mu {\cal G}^{\mu}_{\ c} \right)
     = \left(
  [\mbox{\rm reg}({\cal G})^{\perp}_{\ c}] +
  {\overline{\nabla} }_a {\bf G}^{a}_{\ b} \right)
  \boldsymbol\delta(x^3) \equiv 0
    \ ,
\end{equation}
and vanishes identically due to the Gauss--Codazzi equation
(\ref{G-C}), when we calculate its jump across $S$. Hence,
the Bianchi identity $\nabla_\mu {\cal G}^{\mu}_{\ c}
\equiv 0$ holds universally (in the sense of distributions) for
spacetimes with singular, light-like curvature.

It is worthwhile to notice that the last term in definition
(\ref{Q-fund}) of the tensor density $Q$ of $S$ is identical on
its both sides. Hence, its jump across $S$ vanishes identically.
This way the singular part of the Einstein tensor density
(\ref{Einst-Q}) reduces to:
\begin{equation}\label{Einst-Q-1}
  {{\bf G}^a}_b = [ {Q}^{a}_{\ b} ] = -s
   v_X  \left( [\nabla_b X^a ] - \delta_b^a [\nabla_c X^c ]\right)
     \ .
\end{equation}

\subsection{Energy-momentum tensor of a light-like matter.
Belinfante--Rosenfeld identity} \label{energy-momentum-tensor}

The interaction between a thin
light-like matter-shell and the gravitational field
is described in \cite{jjPRD02}. In particular, all
the properties of such a matter are derived from its Lagrangian density
${L}$, which depends upon (non-specified) matter fields $z^K$ living
on a null-like surface $S$, together with their first derivatives
${z^K}_a:= \partial_a z^K$ and --- of course --- the (degenerate)
metric tensor $g_{ab}$ of $S$:
\begin{equation}\label{L1}
L=L(z^K;{z^K}_a;  g_{ab}) \ .
\end{equation}
We assume that $L$ is an invariant scalar density on $S$.
Similarly as in the standard case of canonical field theory,
invariance of the Lagrangian with respect to reparameterizations of
$S$ implies important properties of the theory: the
Belinfante--Rosenfeld identity and the Noether theorem, which will
be discussed in this Section. To get rid of some technicalities,
we assume in this paper that the matter fields $z^K$ are
``spacetime scalars'', like, e.g.,~material variables of any
thermo-mechanical theory of continuous media (see, e.g.,
 \cite{shell1,KSG}). This means that the Lie derivative
${\cal L}_Y z$ of these fields with respect to a vector field $Y$
on $S$ coincides with the partial derivative:
\[
({\cal L}_Y z)^K = {z^K}_a \ Y^a \ .
\]
The following Lemma characterizes Lagrangians which fulfill the
invariance condition:
\begin{lemma}\label{lem1}
Lagrangian density (\ref{L1}) concentrated on a null hypersurface
$S$ is invariant if and only if it is of the form:
\begin{equation}\label{lagr-form}
L=v_{X} f(z ; {\cal L}_X z ; g)\ ,
\end{equation}
where $X$ is any degeneracy field of the metric $g_{ab}$ on $S$
and $f(\cdot \ ; \cdot \ ; \cdot)$ is a scalar function,
homogeneous of degree 1 with respect to its second variable.
\end{lemma}

\underline{Remark:} Because of the homogeneity of $f$ with respect to
${\cal L}_X z$, the above quantity does not depend upon a choice
of the degeneracy field $X$.

Dynamical properties of such a matter are described by its
canonical energy-momentum tensor density, defined in a standard
way:
\begin{equation}\label{e-m-can}
  {T^a}_b := \frac{\partial { L}}{\partial {z^K}_a}
  {z^K}_b - \delta^a_{\; b} { L} \ .
\end{equation}
It is ``symmetric'' in the following sense:
\begin{theo}\label{th1}
Canonical energy-momentum tensor density ${T^a}_b$ constructed
from an invariant Lagrangian density fulfills identities
(\ref{G-1}) and (\ref{G-2}), i.e.,~the following holds:
\begin{equation}
  {T^a}_b X^b=0\;\;\; {\rm and}\;\;\; T_{ab}=T_{ba}\ .
\end{equation}
\end{theo}

In case of a non-degenerate geometry of $S$, one considers also
the ``symmetric energy-momentum tensor density'' ${\tau}^{ab}$,
defined as follows:
\begin{equation}\label{seym}
{\tau}^{ab}:=2 \frac{\partial L}{\partial g_{ab}} \ .
\end{equation}
In our case the degenerate metric fulfills the constraint: $\det
g_{ab} \equiv 0$. Hence, the above quantity {\em is not} uniquely
defined. However, we may define it, but only {\em up to an
additive term} equal to the annihilator of this constraint. It is
easy to see that the annihilator is of the form $C X^a X^b$.
Hence, the ambiguity in the definition of the symmetric
energy-momentum tensor is precisely equal to the ambiguity in the
definition of $T^{ab}$, if we want to reconstruct it from the well
defined object $T^a{_b}$. This ambiguity is cancelled, when we
lower an index. The next theorem says that for field
configurations satisfying field equations, both the canonical and
the symmetric tensors coincide\footnote{In our convention, the energy
is described by formula: $H={T^0}_0 = {p_K}^0 \dot{z}^K - L \ge
0$, analogous to $H=p \dot{q}-L$ in mechanics and well adapted for
Hamiltonian purposes. This convention differs from the one used in
\cite{Misner}, where the energy is given by $T_{00}$. To keep standard
conventions for Einstein equations, we take standard definition of
the {\em symmetric} energy-momentum tensor ${\tau^a}_b$. This is
why Belinfante--Rosenfeld theorem takes form
${\tau^a}_b=-{T^a}_b$.}. This is an analog of the standard
Belinfante--Rosenfeld identity (see \cite{R-B}). Moreover, Noether
theorem (vanishing of the divergence of $T$) is true. We summarize
these facts in the following:

\begin{theo}\label{ros-bel}
If $L$ is an invariant Lagrangian and if the field configuration
$z^K$ satisfies Euler--Lagrange equations derived from $L$:
\begin{equation}\label{E-L}
  \frac{\partial { L}}{\partial z^K}-\partial_a \frac{\partial
{ L}}{\partial {z^K}_a} = 0 \ ,
\end{equation}
then the following statements are true:
\begin{enumerate}
  \item Belinfante--Rosenfeld identity: canonical
  energy-momentum tensor ${T^a}_b$ coincides with
  (minus --- because of the convention used)
  symmetric energy-mo\-men\-tum tensor ${\tau}^{ab}$:
\begin{equation}\label{R-B}
 {T^a}_b = - {\tau}^{ac}g_{cb} \ ,
\end{equation}
  \item Noether Theorem:
  \begin{equation}\label{Noether}
  {\overline{\nabla}}_a T^a{_b} =0\ .
\end{equation}
\end{enumerate}
\end{theo}
It is shown in \cite{jjPRD02} that
the Einstein equations for the singular part:
\begin{equation}\label{Gtau}
{\bf G}^a{_b} = 8 \pi \tau^a{_b}
\end{equation}
can be derived from an action principle and they contain an
intrinsic part of the Barrab\`es--Israel equations in mixed
(contravariant-covariant) tensor density representation. Let us
notice that if we assume vacuum Einstein equations outside surface
$S$ then, in particular, they imply $\mbox{\rm reg}({\cal
G})^{\perp}_{\ c}=0$ which gives compatibility of
(\ref{Bianchi-fund}) with (\ref{Noether}).
\\[1ex]
\underline{Remark:}
We may also include a regular matter part into the action
and we obtain that the regular part of the energy momentum tensor density
is no longer vanishing.
In that case our null singular matter fulfills the following equation:
\begin{equation}\label{divTsing}
    \mbox{\rm sing}\left(  \nabla_\mu {\cal T}^{\mu}_{\ c} \right)
     = \left(
  [\mbox{\rm reg}({\cal T})^{\perp}{_c}] +
  {\overline{\nabla}}_a {\tau}^{a}{_b} \right)
  \boldsymbol\delta(x^3) = 0 \, ,
\end{equation}
where ${\cal T}_{\mu\nu}$ is the symmetric
energy-momentum tensor density of the
whole matter surrounding our shell $S$.
If $\mbox{\rm reg}({\cal T})^{\mu\nu}$ is derived form the
(regular part of) Lagrangian then eq. (\ref{Gtau}) may be also
considered as a generalized Noether theorem for the the full
(regular + singular) Lagrangian of matter.

\section{Canonical null vector on a single shell}
Let us rewrite the Ricci tensor:
\begin{equation}\label{aric1}
 R_{\mu\nu}=\partial_\lambda{\Gamma}^{\lambda}_{\mu\nu} - \partial_{(\mu}
{\Gamma}^{\lambda}_{\nu )
\lambda}+{\Gamma}^{\lambda}_{\sigma\lambda}
{\Gamma}^{\sigma}_{\mu\nu} - {\Gamma}^{\lambda}_{\mu\sigma}
{\Gamma}^{\sigma}_{\nu\lambda} \ ,
\end{equation}
in terms of the following combinations of Christoffel symbols:
\begin{equation}\label{A-def}
A^{\lambda}_{\mu\nu} := {\Gamma}^{\lambda}_{\mu\nu} -
{\delta}^{\lambda}_{(\mu} {\Gamma}^{\kappa}_{\nu ) \kappa} \ .
\end{equation}
We have:
\begin{equation}
R_{\mu\nu}=\partial_\lambda
A^{\lambda}_{\mu\nu}-A^{\lambda}_{\mu\sigma}
A^{\sigma}_{\nu\lambda} + \frac 13 A^{\lambda}_{\mu\lambda}
A^{\sigma}_{\nu\sigma}.
\end{equation}

The terms quadratic in $A$'s may have only step-like discontinuities.
The derivatives along $S$ are thus bounded and belong to the
regular part of the Ricci tensor. The singular part of the Ricci
tensor is obtained from the transversal derivatives only. In our
adapted coordinate system, where $x^3$ is constant on $S$, we
obtain:
\begin{equation}
{\rm sing}(R_{\mu\nu})=\partial_3
A^3_{\mu\nu}=\boldsymbol\delta(x^3)[A^3_{\mu\nu}] \ ,
\end{equation}
where by $\boldsymbol\delta$ we denote the Dirac
delta-distribution and by square brackets we denote the jump of
the value of the corresponding expression between the two sides of
$S$. Consequently, the singular part of Einstein tensor density
reads:
\begin{equation}\label{Einstein}
{\rm sing}({{\cal G}^\mu}_\nu) := \sqrt{|g|} \  {\rm sing} \left(
{ R^\mu}_\nu - \frac 12 R \right)  =  \boldsymbol\delta(x^3) {{\bf
G}^\mu}_\nu\ ,
\end{equation}
where
\begin{equation}\label{G-grube}
{{\bf G}^\mu}_\nu  :=  \sqrt{|g|} \left(\delta^\beta_\nu g^{\mu
\alpha}  - \frac 12  \delta^\mu_\nu g^{ \alpha \beta} \right)
[A^3_{\alpha\beta}] = [\tQ^{\mu}{_\nu}] \ ,
\end{equation}
\begin{equation} \label{tQ}
{\widetilde  Q}^{\mu\nu} := \sqrt{|g|} \left( g^{\mu \alpha}
g^{\nu \beta} - \frac 12 g^{\mu\nu} g^{ \alpha \beta} \right)
A^3_{\alpha\beta}  \ ,
\end{equation}
 and explicit formulae for
${\widetilde  Q}^{\mu}{_\nu}$ are given in Appendix \ref{GuuA}.
It was also shown in \cite{jjPRD02}
 that the contravariant version of this quantity:
\[ {\rm sing}({\cal G})^{\mu\nu} =
[\tQ^{\mu \nu}]\boldsymbol\delta(x^3) \ ,
\]
is coordinate-dependent and, therefore, does not define any
geometric object.
Let us observe that
 ${\bf G}^{ab}:=[\tQ^{ab}]$
 is not well defined intrinsic tensor density on $S$
 in contrast to ${\bf G}^{a}{_b}=[\tQ^{a}{_b}]$, as
 was shown in Appendix A of \cite{jjPRD02}.
 However, one can extract the following object:
 \begin{equation}\label{Idef}
 I^a := s X^a \frac{{\bf G}^{00}}{X^0\Lambda^0} \, ,
 \end{equation}
 which is well defined because of
  the following 
 \begin{theo}\label{PI}
 The vector field $I$ defined by (\ref{Idef})
 does not depend on the choice of
 the field $X$ and coordinate $x^0$,
  hence it is a well defined intrinsic object on
 the null surface $S$.
 \end{theo}
 \begin{proof}
 Let us express the component ${\bf G}^{00}$ in terms
 of the objects which arise in (1+2+1)-decomposition
 of spacetime (see Appendix \ref{GuuA}):
 \begin{align}
{\bf G}^{00} = &[\tQ^{00}] =  g^{03}[\tQ^0{_3}] + g^{0b}
[\tQ^0{_b}] \nonumber\\ =&
\frac{\lambda}M\left(-[\partial_3\ln\lambda] + m^b [w_b]\right)
- s \left(\frac1{N^2} X^b +\frac sM m^b\right)
\lambda [w_b]  \nonumber\\ = &  - \frac{1}M [\partial_3 \lambda] =
  - s Y^\mu [\partial_\mu \lambda] \ ,
\end{align}
where the last equality holds because tangent to $S$
derivatives $\partial_a \lambda$ are continuous, hence
$[\partial_a \lambda]=0$.
The transformation laws, introduced in \cite{JKC} and
given in Appendix \ref{trr},
 imply that
\[ s X^a \frac{{\bf G}^{00}}{X^0\Lambda^0} =
   - Y^\mu [\partial_\mu \ln \lambda] X^a \]
   is not dependent on the choice of the basis $X,\partial_A,Y$
   at the point $x\in S$. More precisely, for any two tetrads
    $X,\partial_A,Y$ and ${\tilde X},\partial_{\tilde B},{\tilde Y}$
    related by (\ref{tr1})--(\ref{tr2}), (\ref{tr3}) we get
   $[Y(\ln\lambda)]X = [{\tilde Y}(\ln{\tilde\lambda})]{\tilde X}$.
\end{proof}
We also have ${\bf G}{_{\mu\nu}}Y^\mu Y^\nu={\bf G}^{00}$
 because ${\bf G}{_{\mu\nu}}X^\mu=0$ (cf. (\ref{G^perp=0})
 and Appendix \ref{GuuA}).\\[1ex]
\underline{Remark:}
 One can define a symmetric tensor density
 $W := I\otimes\Lambda = \Lambda \otimes I$  on $S$.
 However,  there is no possibility
 to include object $W^{ab}$ into ${\bf G}^{ab}$ unless
 ${\bf G}^a{_b}=0$.
  Moreover, if ${\bf G}^a{_b}$ is vanishing
  (which happens for spherical symmetry cf. Prop. \ref{ssG}), one can check
 from Bianchi identities
 $\nabla_\mu {\cal G}^{\mu}{_\nu} =0$
 that
 \[ \nabla_\mu I^\mu \bigr|_S =0 \]
 for any extension $I^\mu$ which is tangent to $S$.
 Unfortunately, this equation is not intrinsic on $S$.

 The equation (\ref{Gtau}) cannot be completed by
 the equality ${\bf G}^{00}=8\pi\tau^{00}$ on the
 tensor density level because nor ${\bf G}^{ab}$
  neither ${\tau}^{ab}$ are
 geometric objects\footnote{In non-degenerate case
 both tensor densities are well defined.} on $S$. On the other hand,
  the definition
 (\ref{Idef}) allows to complete singular Einstein equations (\ref{Gtau})
 in the following form:
 \be\label{IP}
  I^a = 8\pi P^a \, ,
 \ee
where the vector field $P^a$ defined as follows:
 \be\label{Pdef}
   P^a := s X^a \frac{{\tau}^{00}}{X^0\Lambda^0}
 \ee
contains missing information about singular energy-momentum
tensor density $\tau^{\mu\nu}$.

Let us finish this section with the following observation: for
non-degenerate surface $S$ the tensor density ${\bf G}^{ab}$
(given by (\ref{G-Q})) is well defined. For the null shell $S$ it
splits into two objects: the tensor density ${\bf G}^{a}{_b}$
defined by (\ref{Einst-Q}) and the null vector $I^a$ given by
(\ref{Idef}). This means that the information about the jump of a
``transverse'' extrinsic curvature ${\cal K}_{ab}$ (in
Barrab\`es--Israel approach) is contained in two different
geometric objects -- ${\bf G}^{a}{_b}$ and $I^a$.

\section{Crossing shells}
Let us consider two shells intersecting each other
along surface $\SC$ which is a sphere.
 One can imagine
this situation with the help of Fig. \ref{cs1},
\begin{figure}[h]
\begin{center}
\includegraphics[height=10cm]{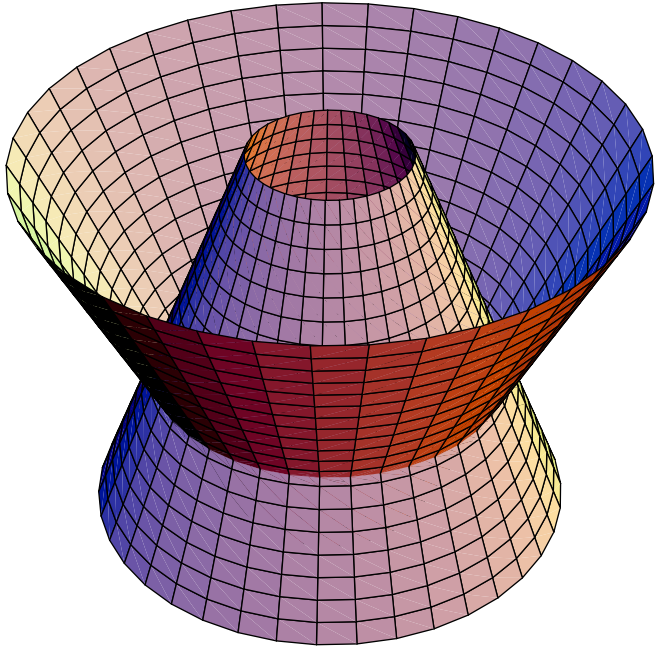}
\end{center}
\caption{Crossing shells}
\label{cs1}
\end{figure}
where one spherical coordinate is suppressed
and the spheres are drawn as one-dimensional circles.

Let us introduce a local coordinate system
$(v,x^A,u)$ around $\SC$, such that
$N_u:=\{ u=u_0 \}$ is the first shell and
$N_v:=\{ v=v_0 \}$ is a second one. Hence
$\SC=N_u \cap N_v$.
The metric takes the form similar to (\ref{gd}) but now
both (transversal to $\SC$) coordinates $u$ and $v$ are null,
i.e. corresponding level three-surfaces are degenerate.
More precisely,
\begin{equation}\label{gdbis}
{g}_{\mu\nu} = \left[
\begin{array}{ccccc} n^A n_A & \vline & n_A
& \vline & sM+m^A n_A \\
 & \vline &  & \vline & \\
\hline & \vline &  & \vline &  \\
 n_A & \vline & g_{AB} & \vline & m_A \\
 & \vline &  & \vline & \\
 \hline & \vline &  & \vline &  \\
 sM+m^A n_A & \vline & m_A & \vline & m^A m_A \\
 \end{array} \right]
\end{equation}
which gives $\sqrt{|\det g_{\mu\nu}|} = \lambda M$,
and the contravariant four-metric takes the form
\begin{equation}\label{gubis}
{g}^{\mu\nu} = \left[ \begin{array}{ccccc}
 0 & \vline &  - s\frac {m^A}{M} & \vline &
\frac {s}{M} \\
 & \vline &  & \vline & \\
\hline & \vline &  & \vline &  \\
  - s\frac {m^A}{M} & \vline &
 {\tilde{\tilde g}}^{AB} + s\frac {n^A m^B + m^A
 n^B}{M} & \vline & - s\frac {n^A}{M} \\
 & \vline &  & \vline & \\
 \hline & \vline &  & \vline &  \\
 \frac s{M} & \vline & - s\frac {n^A}{M} & \vline & 0 \\
 \end{array} \right] \ ,
\end{equation}
where $M > 0$, $s:=\sgn g^{uv}=\pm 1$, $g_{AB}$ is the induced
two-metric on surfaces $\{u={\rm const},\; v={\rm const} \}$
and ${\tilde{\tilde g}}^{AB}$ is its inverse (contravariant)
metric. Both ${\tilde{\tilde g}}^{AB}$ and $g_{AB}$ are used to
rise and lower indices $A,B = 1,2$ of the two-vectors $n^A$ and
$m^A$.

 Let us choose the null vector fields
\[ K:= \partial_v -n^A\partial_A \quad \mbox{and} \quad
   L:= \partial_u -m^A\partial_A \]
 which are tangent to $N_u$ or $N_v$ respectively and
  $g(K,L)=sM$. We can use the coordinates $(v,x^A)$ on the first
 shell $N_u$.
 On the second shell $N_v$ we have the coordinate system $(x^A,u)$.
The canonical vector field $I$ is well defined on both shells:
\be\label{IKL}
I(K)=-\frac{K}M [ L(\ln\lambda)]_u \; ,\quad
 I(L)=-\frac{L}M [ K(\ln\lambda)]_v \; ,
\ee
where the index $u$ or $v$ corresponds to jump across
first or second shell respectively.

Several continuity properties of discontinuities across $\SC$
are implied by the observation that
{\em jump of the jump vanishes} which we explain below
on the example of a real function of two variables.\\
Let $f$ be a function on an open set $U\subset{\mathbb R}^2$
containing point $(0,0)$ such that $f$ is smooth outside
axes (corresponding to our crossing shells),
i.e. $f\in C^k(U')$ for sufficiently large $k\geq 2$
and
 \[ \color{red} U':= U \setminus ( \{(x,y)\in {\mathbb R}^2 \; | \; x=0 \}
\cup \{(x,y)\in {\mathbb R}^2 \; | \; y=0 \}) \; . \]
Moreover, we assume that $f$ is continuous across the axes with finite
jumps of first normal derivatives. More precisely,
the jump \[ \left[\frac{\partial f}{\partial x}\right]_x :=
\lim_{x\rightarrow 0^+} \frac{\partial f}{\partial x}(x,y) -
\lim_{x\rightarrow 0^-} \frac{\partial f}{\partial x}(x,y) \]
is well defined for $y\neq 0$ and splits into upper (positive $y$)
and lower (negative $y$) parts. Under above assumptions we get
\begin{lemma}\label{jjf}
The jump $\left[\frac{\partial f}{\partial x}\right]_x$ is continuous
across $(0,0)$, i.e.
\[ \lim_{y\rightarrow 0^+} \left[\frac{\partial f}{\partial x}\right]_x (y)=
   \lim_{y\rightarrow 0^-} \left[\frac{\partial f}{\partial x}\right]_x (y) \]
\end{lemma}
and the similar property holds on $x$-axis.
\begin{proof}
Let us enumerate the quadrants of the plane:
I, II, III, IV, i.e.
$\mbox{I} \longrightarrow \{(x,y)\in {\mathbb R}^2 \; | \; x > 0, y > 0 \}$,
$\mbox{II} \longrightarrow \{(x,y)\in {\mathbb R}^2 \; | \; x < 0, y > 0 \}$,
$\mbox{III} \longrightarrow \{(x,y)\in {\mathbb R}^2 \; | \; x < 0, y < 0 \}$,
$\mbox{IV} \longrightarrow \{(x,y)\in {\mathbb R}^2 \; | \; x > 0, y < 0 \}$,
and the corresponding restrictions of the function $f$ we denote
by index e.g. the function $f$ in the second quadrant we denote
by $f^{II}$.
Continuity of $f$ and its tangent derivatives across
positive $y$-half-axis implies
$f^I(0,y)=f^{II}(0,y)$ and $\frac{\partial^n f}{\partial y}^{I}(0,y) =
\frac{\partial^n f}{\partial y}^{II}(0,y)$ $n=1,2,\ldots k$, where
the boundary values of $f$ and its derivatives are defined in an
obvious way e.g. $f^I(0,y)=\lim_{x\rightarrow 0^+}f^I(x,y)$.
In particular, we have
\[ \frac{\partial f}{\partial y}^{I}(0,y) =
\frac{\partial f}{\partial y}^{II}(0,y) \quad \mbox{for} \; y>0 \, ,\]
\[ \frac{\partial f}{\partial y}^{IV}(0,y) =
\frac{\partial f}{\partial y}^{III}(0,y) \quad \mbox{for} \; y<0 \, .\]
Passing to the limit at $(0,0)$, we get
\[ \frac{\partial f}{\partial y}^{I}(0,0)
:= \lim_{y\rightarrow 0^+}\frac{\partial f}{\partial y}^{I}(0,y) =
\lim_{y\rightarrow 0^+}\frac{\partial f}{\partial y}^{II}(0,y)
=: \frac{\partial f}{\partial y}^{II}(0,0) \]
and similarly
\[  \frac{\partial f}{\partial y}^{IV}(0,0)
:= \lim_{y\rightarrow 0^-}\frac{\partial f}{\partial y}^{IV}(0,y) =
\lim_{y\rightarrow 0^-}\frac{\partial f}{\partial y}^{III}(0,y)
=: \frac{\partial f}{\partial y}^{III}(0,0) \; . \]
Finally, from the last two equations we get
\[ \frac{\partial f}{\partial y}^{I}(0,0) -
  \frac{\partial f}{\partial y}^{IV}(0,0) =
  \frac{\partial f}{\partial y}^{II}(0,0) -
  \frac{\partial f}{\partial y}^{III}(0,0)
  \]
which implies continuity of
jump $\displaystyle\left[\frac{\partial f}{\partial x}\right]_x $
across $y=0$.
\end{proof}
We can denote symbolically the result as
$\displaystyle\left[ \left[\frac{\partial f}{\partial x}\right]_x\right]_y =0$,
i.e. jump of the jump at the crossing point vanishes.

Using Lemma \ref{jjf} one can show the following
\begin{Theorem}\label{Th1}
The continuity of the metric across null shells implies that
the vector fields $I(K)$ and $I(L)$ are continuous across $\SC$.
\end{Theorem}
Moreover, from Lemma \ref{jjf} we get that ${\bf G}^{a}{_b}(K)$
on $N_u$ and
${\bf G}^{a}{_b}(L)$ on $N_v$ are also continuous\footnote{Although
${\bf G}^{a}{_b}(K)$ does not depend on the choice of the null
field $K$, we keep this argument to distinguish the shells.
Moreover, we should remember that
the coordinates $x^a$ depend on the shell, i.e.
$(x^a)=(v,x^A)$ for $N_u$ but $(x^a)=(u,x^A)$ for $N_v$.}
across $\SC$.
\begin{proof}
From definition (\ref{IKL}) of the null field $K$ and (\ref{gdbis}) we have
\[ I(K) = -\frac{K}{M\lambda} [\partial_u\lambda]_u \; ,\]
hence we apply Lemma \ref{jjf} for the function $\lambda$.
More precisely, we take
\[ f(x,y):= \lambda (u=x+u_0,v=y+v_0, x^A) \]
with fixed coordinates $x^A$, hence the point $x=0,y=0$
corresponds to the fixed
point on $\SC$ with coordinates $x^A$.

For ${\bf G}^{a}{_b}(K)$ we observe that
\[ {\bf G}^{a}{_b}(K)=s\Lambda^a[w_b]=
\frac{\lambda}{2M}K^a K^c[\partial_u g_{cb}]_u \]
which is implied by (\ref{jPl}) and (\ref{jwa})-(\ref{X[w]}).
Moreover, from (\ref{X[w]}) we get $[w_v]=n^A[w_A]$,
hence it is enough to consider
\[ [w_A]=\frac{s}{2M} g_{AB} [\partial_u n^B]_u \]
implied by (\ref{jwa}),
and using Lemma \ref{jjf} for the function $f:=n^B$
we obtain the result.
\end{proof}
 The above Theorem and the considerations
 from Section 2 imply that {\em the dynamics of crossing shells
is described by
equations (\ref{Gtau}) and (\ref{divTsing}) which hold on both
shells plus continuity property across $\SC$}.
\subsection{Spherically symmetric shells}
\begin{theo} \label{ssG}
For spherically symmetric null shell the tensor density
$\; {\bf G}^a{_b}$ is vanishing.
\end{theo}
This implies that the dynamics of the spherical shell is very simple,
i.e. $\tau^a{_b}=0$, hence eqs. (\ref{Gtau})-(\ref{divTsing}) are trivially
satisfied but vector field $I$ is not vanishing as we show in the sequel.
\begin{proof}
From (\ref{Einst-Q-1}), (\ref{jPl})
and (\ref{X[w]}) we get
\begin{equation}\label{Gwa}
  {{\bf G}^a}_b = [ {Q}^{a}_{\ b} ] = \Lambda^a[w_b]
\end{equation}
but spherical symmetry  gives
$[w_A]=0$ and, moreover, (\ref{X[w]}) implies $[w_0]=0$.
\end{proof}
Let us check the value of $I$ for the
spherical null shell which arises from matching
two Schwarzschild metrics along spherically symmetric
null surface.
\be\label{gs}
g_i= -\left(1-\frac{2m_i}{r_i}\right)\rd u^2 -2\rd u \rd r_i
+r_i^2 \rd\Omega \; ,
\quad i=1,2 \, ,
\ee
where
\[ \rd\Omega:=\rd\theta^2 + \sin^2\theta\rd\varphi^2\; . \]
We take $u\geq 0$ for $g_1$ and $u\leq 0$ for $g_2$,
and $r_i(R,u):= R + \frac{m_i}r u$. This implies that
the full metric is continuous in coordinates $(u,R)$ across the shell $u=0$.
More precisely,
\[ g_1\bigr|_{u=0} =
-\left(1-\frac{2m_1}R\right)\rd u^2 -2\rd u \left( \rd R +
\frac{m_1}R \rd u \right) + R^2 \rd\Omega =
-\rd u^2 -2\rd u  \rd R
+ R^2 \rd\Omega = \]
\[ = -\left(1-\frac{2m_2}R\right)\rd u^2 -2\rd u \left( \rd R +
\frac{m_2}R \rd u \right) + R^2 \rd\Omega =
g_2\bigr|_{u=0} \; .
 \]
 Moreover, if we choose null field $X=\frac{\partial}{\partial R}$
 then the transversal field
 may be chosen as
 $Y=-\frac{\partial}{\partial u}$ and $\lambda=r_i^2\sin\theta$,
 hence
 \[ Y(\ln\lambda)\bigr|_{u=0}= -2\frac{\partial}{\partial u}\ln \left(R +
 \frac{m_i}R u\right)\bigr|_{u=0}= -\frac{2m_i}{R^2} \]
and finally
\be\label{I12} I = 2\frac{m_1-m_2}{R^2} X \, . \ee
Next, for crossing two spherical null shells we may check
the Dray--t'Hooft--Redmount formula \cite{DH}, \cite{Red} as follows:
firstly we apply Theorem \ref{Th1} which from continuity
of the metric implies continuity of the vector field $I$,
secondly we check that
the vector field $I$ is continuous through the crossing
sphere iff the Dray--t'Hooft--Redmount formula
is true.
\begin{Theorem}
If the shells are spherically symmetric than
continuity of the
vector field $I$ gives Dray--t'Hooft--Redmount formula
(\ref{DHR}).
\end{Theorem}
\begin{proof}
Let us consider the full description of
crossing spherically symmetric null shells which
can be nicely given in Kruskal--Szekeres coordinates (instead of
Eddington--Finkelstein used in (\ref{gs})). We assume four domains
(cf. Fig. \ref{rys1}) equipped with the Schwarzschild metrics
\be\label{KSS}
 g_i = -\frac{32m_i^3}{r_i}\exp\left(-\frac{r}{2m_i}\right)\rd u_i \rd v_i
 + r_i^2 \rd\Omega \; , \quad i=1,2,3,4 \, ,
\ee
where
 $r_i = 2m_i\kappa(-u_i v_i)$ and
 the Kruskal function $\kappa$ is defined by its inverse
$\kappa^{-1}(x) = (x-1)e^x$
on the interval $(0,\infty)\subset {\mathbb R}$.
One can easily check the following identity for the first
derivative of $\kappa$:
\be\label{derkap} \kappa'= \frac{\exp(-\kappa)}{\kappa} \, . \ee
The four domains $M_i$ ($i=1,2,3,4$)
are matched together along null surfaces
$\{x\in M_i \; | \; u_i=\alpha_i \}\subset M_i$ and
  $\{x\in M_i \; | \; v_i=\beta_i \}\subset M_i$,
as is shown on Fig. \ref{rys1}.
\begin{figure}
\begin{center}
\begin{picture}(150,150)(0,0)
\thicklines
\put(0,0){\line(1,1){150}}
\put(150,0){\line(-1,1){150}}

\put(110,75){\makebox(0,0)[l]{$u_4\leq\alpha_4$, $v_4\geq\beta_4$}}
\put(75,140){\makebox(0,0)[c]{$u_1\geq\alpha_1$, $v_1\geq\beta_1$}}
\put(40,75){\makebox(0,0)[r]{$u_2\geq\alpha_2$, $v_2\leq\beta_2$}}
\put(75,10){\makebox(0,0)[c]{$u_3\leq\alpha_3$, $v_3\leq\beta_3$}}

\put(85,75){\makebox(0,0)[l]{4}}
\put(75,85){\makebox(0,0)[c]{1}}
\put(65,75){\makebox(0,0)[r]{2}}
\put(75,65){\makebox(0,0)[c]{3}}
\thinlines

\put(-20,7){\line(0,1){3}}
\put(-15,10){\oval(10,10)[tl]}
\put(-15,15){\vector(1,0){25}}
\put(-15,0){\makebox(0,0)[r]{first shell $N_{23}$}}

\put(-20,127){\line(0,1){3}}
\put(-15,130){\oval(10,10)[tl]}
\put(-15,135){\vector(1,0){25}}
\put(-15,120){\makebox(0,0)[r]{second shell $N_{12}$}}

\put(169,7){\line(0,1){3}}
\put(164,10){\oval(10,10)[tr]}
\put(164,15){\vector(-1,0){25}}
\put(164,0){\makebox(0,0)[l]{second shell $N_{34}$}}

\put(169,127){\line(0,1){3}}
\put(164,130){\oval(10,10)[tr]}
\put(164,135){\vector(-1,0){25}}
\put(164,120){\makebox(0,0)[l]{first shell $N_{14}$}}

\put(22,45){\line(0,-1){5}}
\put(27,40){\oval(10,10)[bl]}
\put(27,35){\vector(1,0){5}}
\put(29,52){\makebox(0,0)[r]{$u_2=\alpha_2$}}

\put(108,114){\vector(0,-1){3}}
\put(103,114){\oval(10,10)[tr]}
\put(101,119){\line(1,0){3}}
\put(98,119){\makebox(0,0)[cr]{$u_1=\alpha_1$}}

\put(108,102){\vector(0,1){3}}
\put(113,102){\oval(10,10)[bl]}
\put(111,97){\line(1,0){3}}
\put(118,95){\makebox(0,0)[cl]{$u_4=\alpha_4$}}

\put(38,32){\vector(0,1){3}}
\put(43,32){\oval(10,10)[bl]}
\put(41,27){\line(1,0){3}}
\put(48,25){\makebox(0,0)[cl]{$u_3=\alpha_3$}}

\end{picture}
\end{center}
\caption{Matching domains}
\label{rys1}
\end{figure}
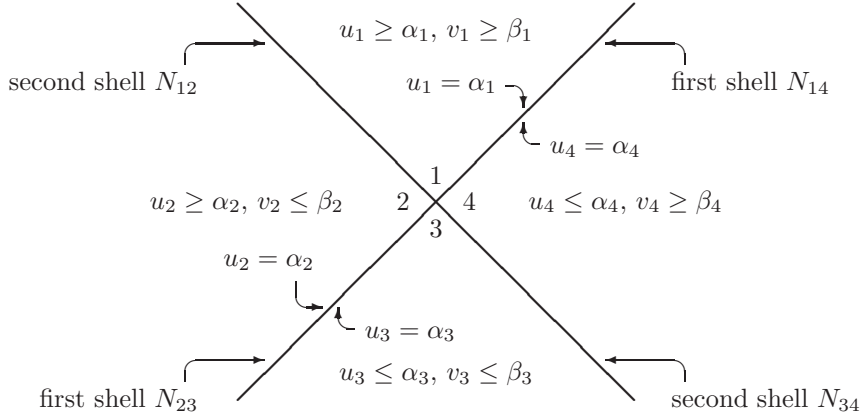

The coordinates $v_1$, $v_4$  on the shell $N_{14}$ do not match
 but
\be\label{rv14} r=2m_1\kappa(-\alpha_1 v_1)= 2m_4\kappa(-\alpha_4 v_4) \ee
is the same on both sides and can be chosen as a
coordinate on the surface $N_{14}$. This equality also means that
$\lambda$ is continuous across this shell.
On the other hand the continuity of the term
$\frac{32m_i^3}{r_i}\exp\left(-\frac{r}{2m_i}\right)\rd u_i \rd v_i$
across $N_{14}$ implies
\[ \frac{m_1^3}{r}\exp\left(-\frac{r}{2m_1}\right)\rd u_1 \rd v_1
   \Bigr|_{u_1=\alpha_1}
   = \frac{m_4^3}{r}\exp\left(-\frac{r}{2m_4}\right)\rd u_4 \rd v_4
    \Bigr|_{u_4=\alpha_4} \; ,
   \]
hence using (\ref{derkap}) and (\ref{rv14}) we obtain the transformation law
 between first derivatives of coordinates $u_1$ and $u_4$:
\be\label{u14}
\frac{du_4}{du_1}  = \left(\frac{m_1}{m_4}\right)^3
\exp\left(-\frac{r}{2m_1}\right)
\exp\left(\frac{r}{2m_4}\right) \frac{dv_1}{dv_4} =
\frac{m_1\alpha_4}{m_4\alpha_1} \; .
\ee
Moreover, the null vector field $X$ tangent to the
first shell can be represented in $M_1$ as follows:
\[ X=\frac{\partial}{\partial r}= \left(\frac{d r}{d v_1}\right)^{-1}
\frac{\partial}{\partial v_1} \; , \]
and using (\ref{derkap}) we have
\[ \frac{d r}{d v_1}= -2m_1\alpha_1\kappa'(-\alpha_1 v_1)=
   -2m_1\alpha_1 \left( \kappa(-\alpha_1 v_1)
   \exp[\kappa(-\alpha_1 v_1)]\right)^{-1} \; ,\]
hence
\[ X=
-\frac{\kappa(-\alpha_1 v_1)\exp(\kappa(-\alpha_1 v_1))}{2\alpha_1 m_1}
\frac{\partial}{\partial v_1} \; . \]
The transversal vector field
\[ Y= \frac{\alpha_1}{4m_1}\frac{\partial}{\partial u_1} \]
fulfills  normalization condition $g_1(X,Y)=1$.
Moreover, using equality
\[ Y(\ln\lambda)= \frac{\alpha_1}{2m_1r_1}\frac{\partial r_1}{\partial u_1}
= \frac{r_1-2m_1}{r_1^2} \]
and the similar one in $M_4$
we can check the formula (\ref{I12})
in new coordinate representation
\[ I_{14} = - [Y(\ln\lambda)]X = \frac{2m_1-2m_4}{r^2} X \, .\]
Similar considerations for the first shell $N_{23}$ give the following
expression for the vector field (\ref{Idef}):
\[ I_{23} = \frac{2m_2-2m_3}{r^2} X \, ,\]
where now $r=2m_2\kappa(-\alpha_2 v_2)= 2m_3\kappa(-\alpha_3 v_3)$
and
\[ X=\frac{\partial}{\partial r}=
-\frac{\kappa(-\alpha_2 v_2)\exp(\kappa(-\alpha_2 v_2))}{2\alpha_2 m_2}
\frac{\partial}{\partial v_2} =
-\frac{\kappa(-\alpha_3 v_3)\exp(\kappa(-\alpha_3 v_3))}{2\alpha_3 m_3}
\frac{\partial}{\partial v_3} \, .
  \]
We can compare $I_{14}$ with $I_{23}$ across $\SC$
by using the transformation law (cf. (\ref{u14}))
between $v_4$ and $v_3$
\be\label{v34}
 \frac{d v_4}{d v_3} = \frac{\beta_4 m_3}{\beta_3 m_4} \ee
which is implied by continuity of the metrics
$g_3$ and $g_4$ across
second shell $N_{34}$ ($v_3=\beta_3$ and $v_4=\beta_4$).
Finally, we obtain
\begin{eqnarray}
I_{23}(v_3=\beta_3) & = & \nonumber - \frac{2(m_2-m_3)}{r^2}
\frac{\kappa(-\alpha_3 \beta_3)
\exp(\kappa(-\alpha_3 \beta_3))}{2\alpha_3 m_3}
\frac{\partial}{\partial v_3} \\ & = & \nonumber
 - \frac{2(m_2-m_3)}{r^2}
\frac{\frac{r}{2m_3}
\exp(\frac{r}{2m_3})}{2\alpha_3 m_3}
\frac{\beta_4 m_3}{\beta_3 m_4}
\frac{\partial}{\partial v_4} \\ & = & \nonumber
-\frac{(m_2-m_3)\beta_4\exp(\frac{r}{2m_3})}{2r\alpha_3 \beta_3 m_4 m_3}
\frac{\partial}{\partial v_4}
 \end{eqnarray}
and
\[ I_{14}(v_4=\beta_4) = -\frac{2(m_1-m_4)}{r^2}
\frac{\frac{r}{2m_4}
\exp(\frac{r}{2m_4})}{2\alpha_4 m_4}
\frac{\partial}{\partial v_4} \, ,
\]
hence $I_{23}=I_{14}$ on $\SC$ implies
\[ \frac{(m_1-m_4)\exp(\frac{r}{2m_4})}{2r\alpha_4 m_4} =
 \frac{(m_2-m_3)\beta_4\exp(\frac{r}{2m_3})}{2r\alpha_3 \beta_3 m_3}
 \, ,
\]
or
\be\label{m1234} (m_1-m_4)\alpha_3 \beta_3 m_3 \exp(-\frac{r}{2m_3})=
 (m_2-m_3)\alpha_4 \beta_4 m_4 \exp(-\frac{r}{2m_4}) \, . \ee
Moreover, on $\SC$
\[ \alpha_i\beta_i\exp(-\frac{r}{2m_i})=1-\kappa(-\alpha_i\beta_i)=
   1-\frac{r}{2m_i} \]
 which applied to (\ref{m1234}) gives
\[ (m_1-m_4)(r-2m_3)= (m_2-m_3)(r-2m_4) \, , \]
which is equivalent to Dray--t'Hooft--Redmount formula
\be\label{DHR} (r-2m_1)(r-2m_3)= (r-2m_2)(r-2m_4) \, . \ee
\end{proof}
In the above proof we have restricted ourselves  to the case
of positive masses $m_i$ and to the matching null surfaces
which are not horizons. The analysis of possible
special cases one can find in \cite{H-Kou1} but
obviously the formula (\ref{DHR}) remains valid for any special case.

\section*{Acknowledgments}
The author is much indebted to Petr H\'{a}j\'{\i}\v{c}ek for
inspiring discussions. This work was partially
supported by Swiss National Funds.

\appendix
\section{Transformation rules}
\label{trr}
The triad $(X, \partial_A )$ on $S$ depends upon a particular
$(2+1)$-decomposition of $S$, given by the choice of the time
coordinate $x^0$ on $S$. However, several objects
constructed by means of the triad do not depend upon this choice
and describe the geometry of $S$. To prove this independence,
observe that we have the following transformation law:
\begin{eqnarray}
{\tilde X} & = & c X \ , \label{tr1}
\\
{\tilde \partial}_{\tilde B} & = &   C_{\tilde B}^{\ A} \partial_A
+f_{{\tilde B}} X
 \ ,
 \label{tr2}
\end{eqnarray}
 where $({\tilde X},{\tilde \partial}_{\tilde B})$ is the new
 triad, corresponding to the new coordinate system
 $(\tilde{x}^{\tilde{a}})$ on $S$.
The coefficient $c$ may be obtained from the following equation:
\begin{eqnarray}\label{c}
1=\langle d\tilde{x}^0 , \tilde{X}\rangle &=& \langle
\frac{\partial \tilde{x}^0}{\partial x^A} dx^A+\frac{\partial
\tilde{x}^0}{\partial x^0}dx^0 , c X \rangle \nonumber\\
&=&c\left(-\frac{\partial \tilde{x}^0}{\partial
x^A}n^A+\frac{\partial \tilde{x}^0}{\partial x^0}\right) \ ,
\end{eqnarray}
hence,
\begin{equation}
c=\left(\frac{\partial \tilde{x}^0}{\partial x^0}-\frac{\partial
\tilde{x}^0}{\partial x^A}n^A\right)^{-1}\ .
\end{equation}
 On the other hand, we have:
\begin{eqnarray}
\partial_{\tilde{B}}&=&\frac{\partial x^A}{\partial
\tilde{x}^{\tilde{B}}}\partial_A+\frac{\partial x^0}{\partial
\tilde{x}^{\tilde{B}}}\left(X+n^A\partial_A\right)\nonumber\\
&=&\left(\frac{\partial x^A}{\partial
\tilde{x}^{\tilde{B}}}+\frac{\partial x^0}{\partial
\tilde{x}^{\tilde{B}}}n^A\right)\partial_A+\frac{\partial
x^0}{\partial \tilde{x}^{\tilde{B}}}X\ ,
\end{eqnarray}
hence,
\begin{eqnarray}
C_{\tilde{B}}^{\ A}&=&\frac{\partial x^A}{\partial
\tilde{x}^{\tilde{B}}}+\frac{\partial x^0}{\partial
\tilde{x}^{\tilde{B}}}n^A\ ,\\ f_{\tilde{B}}&=&\frac{\partial
x^0}{\partial \tilde{x}^{\tilde{B}}}\ . \label{ef}
\end{eqnarray}
 The
transformation law for $g_{AB}$:
\begin{equation}
g_{\tilde{A}\tilde{B}}=C_{\tilde{A}}^{\ A}C_{\tilde{B}}^{\ B}
g\left(\partial_A +f_{A}X,\partial_B+f_{B}
X\right)=C_{\tilde{A}}^{\ A}C_{\tilde{B}}^{\ B}  g_{AB}
\end{equation}
implies:
\begin{equation}
\tilde{\lambda}=\lambda\det C_{\tilde{A}}^{\ B} \ .
\end{equation}
In order to complete the triad $(X, \partial_A)$ on $S$ to a
tetrad in $M$ it is useful to choose a transverse field $Y$
fulfilling the following ``normalization conditions'':
\begin{eqnarray}\label{normXY}
g(Y,X) & = & 1 \ ,
\\ \label{normYA}
g(Y,\partial_A) & = & 0 \ .
\end{eqnarray}
These equations {\em do not} determine $Y$ uniquely, but {\em
modulo} an additive term proportional to $X$: a ``gauge
transformation''
\begin{equation}\label{YtoY+hX}
  Y \rightarrow Y+ hX \ ,
\end{equation}
with an arbitrary scalar field $h$ is always possible. Extending
coordinate $x^0$ from $S$ to a neighbourhood of $S$, we may choose the
following transverse field:
\begin{equation}\label{igrek}
Y=\frac sM \left( \partial_3-m^A \partial_A \right) \ .
\end{equation}
We stress, however, that this particular choice of $Y$
 depends not only upon a
$(2+1)$-decomposition of $S$, but also on a $(3+1)$-decomposition
of $M$ in a neighbourhood of $S$. Because of
(\ref{normYA}), the vectors $X$ and $Y$ span the bundle of vectors
normal to $S$.

The transformation law for $Y$,
when passing from one to another $(2+1)$-decomposition of $S$,
reads:
\begin{eqnarray}
{\tilde Y} & = & c^{-1} \left( Y - k^A\partial_A \right)  + hX \ ,
 \label{tr3}
\end{eqnarray}
where the scalar field $h$ is arbitrary (it is determined by the
extension of the $(2+1)$-decomposition of $S$ to a
$(3+1)$-decomposition of $M$), and the coefficients $k^A$ are
uniquely determined by equation
\begin{equation}\label{ka}
  f_{{\tilde B}} =  C_{\tilde B}^{\ A} g_{AC}k^C \ ,
\end{equation}
with $f_{{\tilde B}}$ given by (\ref{ef}). Despite of the freedom
in choice of $Y$, some geometric objects constructed with help of
the tetrad $(X, \partial_A , Y)$ do not depend upon this choice
and characterize only the geometry of $S \subset M$.

\section{Structure of the singular Einstein tensor}
\label{GuuA}
We are going to relate the
coordinate-dependent quantity $\tQ^{\mu \nu}$ with the external
curvature $Q^a{_b}$ of $S$. We use the form of the metric
introduced in \cite{JKC}:
\begin{equation}\label{gd}
{g}_{\mu\nu} = \left[
\begin{array}{ccccc} n^A n_A & \vline & n_A
& \vline & sM+m^A n_A \\
 & \vline &  & \vline & \\
\hline & \vline &  & \vline &  \\
 n_A & \vline & g_{AB} & \vline & m_A \\
 & \vline &  & \vline & \\
 \hline & \vline &  & \vline &  \\
 sM+m^A n_A & \vline & m_A & \vline &
\left( \frac{M}{N}\right)^2+m^A m_A \\
 \end{array} \right] \ ,
\end{equation}
and
\begin{equation}\label{gu}
{g}^{\mu\nu} = \left[ \begin{array}{ccccc} - \left( \frac 1N
\right)^2 & \vline & \frac {n^A}{N^2} - s\frac {m^A}{M} & \vline &
\frac {s}{M} \\
 & \vline &  & \vline & \\
\hline & \vline &  & \vline &  \\
 \frac {n^A}{N^2} - s\frac {m^A}{M} & \vline &
 {\tilde{\tilde g}}^{AB} - \frac {n^A n^B}{N^2} + s\frac {n^A m^B + m^A
 n^B}{M} & \vline & - s\frac {n^A}{M} \\
 & \vline &  & \vline & \\
 \hline & \vline &  & \vline &  \\
 \frac s{M} & \vline & - s\frac {n^A}{M} & \vline & 0 \\
 \end{array} \right] \ ,
\end{equation}
where $M > 0$, $s:=\sgn g^{03}=\pm 1$, $g_{AB}$ is the induced
two-metric on surfaces $\{x^0={\rm const},\; x^3={\rm const} \}$
and ${\tilde{\tilde g}}^{AB}$ is its inverse (contravariant)
metric. Both ${\tilde{\tilde g}}^{AB}$ and $g_{AB}$ are used to
rise and lower indices $A,B = 1,2$ of the two-vectors $n^A$ and
$m^A$.

Formula (\ref{gd}) implies: $\sqrt{|\det g_{\mu\nu}|} = \lambda
M$. Moreover, the object $\Lambda^a$ defined by formula
(\ref{Lambda}), takes the form $\Lambda^a=\lambda X^a$, where
$\lambda$ is given by formula (\ref{lambda}) and $X:=\partial_0 -
n^A\partial_A$. This means that we have chosen the following
degeneracy field: $X^\mu = (1, -n^A , 0)$.

For calculational purposes it is useful to rewrite the
two-dimensional inverse metric ${\tilde{\tilde g}}^{AB}$ in
three-dimensional notation, putting ${\tilde{\tilde g}}^{0a} :=
0$. This object satisfies the obvious identity:
\[
 {\tilde{\tilde g}}^{ac} g_{cb}=\delta^a{_b}-X^a\delta^0{_b} \, .
\]
Hence, the contravariant metric (\ref{gu}) may be rewritten as
follows:
\begin{equation}\label{g2i}
  g^{ab} = {\tilde{\tilde g}}^{ab} -\frac1{N^2} X^aX^b -
 \frac{s}M ( m^a X^b + m^b X^a ) \ ,
\end{equation}
where $m^a:= {\tilde{\tilde g}}^{aB}m_B$, so that $m^0:=0$, and
\[ g^{3\mu} = \frac sM X^\mu \ . \]
It may be easily checked (see, e.g.,~\cite{JKC}, page 406) that
covariant derivatives of the field $X$ {\em along} $S$ are equal
to:
\begin{equation} \label{gradX}
  \nabla_a X = -w_a X - l_{ab}{\tilde{\tilde g}}^{bc}\partial_c
  \  ,
\end{equation}
where
\begin{equation}\label{wu-def}
w_a  := - X^\mu \Gamma^0_{\mu a} \ ,
\end{equation}
and
\begin{equation}\label{el-def}
l_{ab}:=  -g(\partial_b,\nabla_a X) = g(\nabla_a\partial_b, X) =
X_\mu \Gamma^\mu_{ab} \ .
\end{equation}

Moreover,
\begin{align}\label{dLambda}
  \partial_c \Lambda^c  = &
  -\lambda g^{ab} l_{ab} =
-\lambda {\tilde{\tilde g}}^{ab} l_{ab} = -\lambda  l \ ,
\end{align}
where $l = {\tilde{\tilde g}}^{ab} l_{ab}$.

The following lemma was proved in \cite{jjPRD02}:
\begin{lemma}
The object ${\tQ}^a{_b}$ is related to $Q^a{_b}$ as follows:
\begin{equation}\label{PQ}
 s {\tQ}^a{_b} = s Q^a{_b} -\frac12 \lambda l \delta^a{_b}+ \Lambda^a
\abar_b -\delta^a{_b} \Lambda^c \abar_c \ ,
\end{equation}
where $\displaystyle \abar_c := \frac12 \partial_c \ln \left(\frac
M\lambda \right)$.
\end{lemma}
{}Moreover, from definition (\ref{Q-fund}) and property (\ref{gradX}) one
can check that
\begin{eqnarray}\label{Qli}
 sQ^a{_b} & = & \lambda\delta^a{_b}\nabla_c X^c -\lambda \nabla_b X^a -
 \delta^a{_b} \partial_c \Lambda^c \nonumber\\
 &=& -\lambda\delta^a{_b}(w_cX^c+l) + \lambda (w_bX^a+ {\tilde{\tilde g}}^{ac}
 l_{cb}) + \delta^a{_b} \lambda l \nonumber\\
 &=& \lambda  {\tilde{\tilde g}}^{ac} l_{cb}  +
    \Lambda^a w_b -\delta^a{_b} \Lambda^c w_c \, .
\end{eqnarray}
\underline{Remark:} Formula (\ref{Qli}), together with
$l_{ab}X^b=0=g_{ab}X^b$, gives us the orthogonality condition
$Q^a{_b}X^b =0$ and symmetry of the tensor
$Q_{ab}:=g_{ac}Q^c{_b}$.

Now, we would like to examine the properties of ${\bf G}^{\mu\nu}
= [\tQ^{\mu \nu}]$. {}From continuity of the metric across $S$ we
obtain
\begin{equation}\label{jlab} [l_{ab}] = s M [A^3_{ab}] =
 s M [\Gamma^3_{ab}] = X^c [\Gamma_{cab}]
 = 0 \, ,
\end{equation}
\begin{align}\label{jPl}
s [{\tQ}^a{_b}] &=  \Lambda^a [A^3_{3b}] - \delta^a{_b} \Lambda^c
[A^3_{3c}] \nonumber\\& = \Lambda^a [w_{b}] - \delta^a{_b}
\Lambda^c [w_{c}] = s[Q^a{_b}] \ ,
\end{align}
and
\begin{equation}\label{jP3mu} [{\tQ}^3{_\mu}] = 0 \
\end{equation}
because $s {\tQ}^3{_3} = -\frac12 \lambda l$
  and $s {\tQ}^3{_a} = 0$.

Finally, the missing component $[\tQ^a{_3}]$ has the following form:
\begin{equation} \label{jQa3}
 [\tQ^a{_3}] =   s\Lambda^a \left\{ -[\partial_3\ln\lambda] +m^b [w_b]
\right\} + M \lambda {\tilde{\tilde g}}^{ab}  [w_b] \, .
\end{equation}
We also have from
\begin{equation} \label{jwa} [w_a] =  - X^b g^{03}[\Gamma_{3\, ba}]
 =\frac{s}{2M}  X^b [ g_{ab,3}]
\end{equation}
that
\begin{equation}\label{X[w]}
  X^a[w_a] = \frac{s}{2M} [ X^a X^b g_{ab,3}] = 0 \, .
\end{equation}

Using these results from (\ref{jP3mu}) one can
easily check the property (\ref{G^perp=0})

\begin{eqnarray*}
{\bf G}^{33}=[\tQ^{33}] & = & g^{33}[\tQ^3{_3}] + g^{3b}
[\tQ^3{_b}] = 0 \ , \\ {\bf G}^{3a}=[\tQ^{3a}] & = &
g^{33}[\tQ^a{_3}] + g^{3b} [\tQ^a{_b}]
 = -\frac sM [ X^b Q^a{_b}] = 0 \ ,
\end{eqnarray*}
where we used the equality $[\tQ^a{_b}] =  [Q^a{_b}]$ which is
crucial to admit that the object ${\bf G}^a{_b}$ is a well defined
geometric object on $S$.

\section{Gauss--Codazzi equations}
It was shown in \cite{jjPRD02} that
\begin{align}\label{fund2}
s{\cal G}^{3}_{\ a}
 =& -s\partial_b {Q}^{b}{_a} +\frac 12 s{Q}^{bc}
g_{bc , a} +\lambda \partial_a l \ ,
\end{align}
where we have used the formula
\[
    sQ^{ab}= \lambda  {\tilde{\tilde g}}^{ac}
    {\tilde{\tilde g}}^{bd} l_{cd} +
    ( \Lambda^a {\tilde{\tilde g}}^{bc} +
    \Lambda^b {\tilde{\tilde g}}^{ac}
    - {\tilde{\tilde g}}^{ab} \Lambda^c) w_c \, .
\]


\end{document}